# Classifying Human Activities with Inertial Sensors: A Machine Learning Approach

Hamza Ali Imran
Department of Computing
School of Electrical Engineering &
Computer Science,
National University of Science and
Technology
Islamabad, Pakistan
himran.mscs18seecs@seecs.edu.pk

Saad Wazir
Department of Computing
School of Electrical Engineering &
Computer Science,
National University of Science and
Technology
Islamabad, Pakistan
swazir.mscs18seecs@seecs.edu.pk

Usman Iftikhar
Department of Computing
School of Electrical Engineering &
Computer Science,
National University of Science and
Technology
Islamabad, Pakistan
uiftikhar.mscs18seecs@seecs.edu.pk

Usama Latif
Specialist VAS and HLR
Pakistan Mobile Communications
Limited
usamalatif417@gmail.com

*Abstract*—Human Activity Recognition (HAR) is an ongoing research topic. It has applications in medical support, sports, fitness, social networking, human-computer interfaces, senior care, entertainment, surveillance, and the list goes on. Traditionally, computer vision methods were employed for HAR, which has numerous problems such as secrecy or privacy, the influence of environmental factors, less mobility, higher running costs, occlusion, and so on. A new trend in the use of sensors, especially inertial sensors, has lately emerged. There are several advantages of employing sensor data as an alternative to traditional computer vision algorithms. Many of the limitations of computer vision algorithms have been documented in the literature, including research on Deep Neural Network (DNN) and Machine Learning (ML) approaches for activity categorization utilizing sensor data. We examined and analyzed different Machine Learning and Deep Learning approaches for Human Activity Recognition using inertial sensor data of smartphones. In order to identify which approach is best suited for this application.

## I. INTRODUCTION

Human Activity Recognition (HAR) is a current research topic. Its uses include healthcare, sports, education, social networking, environmental support, senior care, entertainment, child monitoring, and the list goes on [1-4]. Human Activity recognition is one of the goals of the Body Area Network, which is an extension of the Wireless Sensor Network [5]. Computer Vision approaches have historically been employed for HAR [6-7], which has a number of drawbacks, including secrecy, environmental impact, less portability, increased operating expenses, occlusion, and so on.

A new trend in the use of sensors, particularly inertial sensors, has lately arisen [8-9]. There are numerous advantages to employing sensor data rather than typical computer vision approaches for HAR. The use of sensor data addresses nearly all of the visual technique's drawbacks. Different types of sensors, including accelerometers, gyroscopes, and heart rate monitors, are utilized in HAR. There has been a lot of research done on the use of Deep Neural Networks (DNN) and ML approaches for HAR.

In this study, we explored the use of a variety of ML algorithms, including the K Nearest Neighbors Classifier, Support Vector Machine (SVM) with various kernels, Multi-Layer Perceptron, and Naive Bayesian Classifier. Feature engineered data was utilized to train ML classifiers. Accuracy was regarded as the performance metric. The dataset was separated into 70 percent Training, 15% Validation, and 15% Test sets. Our findings suggest that SVM with linear kernel performs the best for HAR using smartphone inertial sensors data.

Section II contains a review of the literature, Section III discusses the dataset used for experimentation, Section IV discusses the details of the Machine Learning classifiers that are being trained, Section V compares the results of all the models being trained, and Section VI concludes the paper.

## II. RELATED WORK

H. A. Imran et. al. presented a novel model convolutional neural network [1] which has inception-like multi-sized kernel models and dense likes between those models for HAR using inertial sensors data of smartphones. They have evaluated their model on the UCI-HAR/SMARTPHONE dataset which we have used for this study and they have achieved 89.44 % accuracy.

K. Mehmood et. al. presented a convolutional neural network for HAR using raw sensors data of initial sensors of UCI-HAR/SMARTPHONE [2]. They have also evaluated their model on a smartphone dataset and have achieved 91.61 % accuracy.

C. Xu, D et al. built a Deep Neural Network called InnoHAR in [3]. It is built on a foundation of Inception Neural Network and GRU layers. They ran trials on three separate datasets: the OPPORTUNITY Dataset, the PAMAP2 Datasets, and the SMARTPHONEDataset.

A one-dimensional convolution neural network for human activity recognition has been proposed by researchers in [8]. They developed a magnitude vector from triaxial accelerometer data (x,y,z) from a smartphone. They compared their design to the Random Forest Classifier. They had a 92.71 percent accuracy rate. The dataset only had three classes: walk, run, and still.

[9] In order to evaluate human operations, researchers used portable inertial sensors. The sensor data are utilized featured engineered. Linear Discriminant Analysis (LDA) and Principal Kernel Component Analysis (KPCA) were employed for robust features. A Deep Belief Neural Network (DBN) was trained on robust features. The findings were compared using SVM and ANN. They utilized a dataset of 12 classifications, which was open to the public. They attained an overall accuracy of 95,85% for the suggested DBN, 89.06% for the ANN, and 94.12% for the SVM.

In [10] researchers have been able to identify events by means of modified accelerometer information. They directly transform input signals to a reference coordinate system based on the gyroscope orientation and orientation detector's rotation matrix (Euler angle conversion). The advantage of the conversion is that the behavior and detection may be determined and that the sensors remain non-partisan. The class number was 5. A smartphone was used for data collection. The findings indicate 84.77% of normal orientation unbiased accuracy, 17.26% better than the ones without input processing.

The authors of [11] have studied numerous tools and approaches, including multiple machine learning algorithms and neural network methods, that may be used to detect human activity. The study contains different observations based on the environment, the type of facts utilized together with statistics on the accelerometer, other sensor data, sensor placing, and implementation techniques. Based on these settings and on the concept of computational difficulties, various techniques are contrasted. Finally, issues are also explored in HAR. They concluded that no one approach, great for recognizing any action, is intended to take into account numerous aspects in order to choose a preferred approach for the desired application and thus the approach. So, in spite of getting several strategies, some of the challenges nonetheless continue to be open and need to be resolved.

To increase HAR performance, the author of the article [12] suggests adopting a dividing and conquering technique for 1D CNN-based HAR methods that use sharpening of test data. By utilizing an uncertainty matrix to categorize abstract sports, they created an a-level HAR approach. The employment of a Gaussian filter to sharpen test data has resulted in a wide range of possible increases in activity detection accuracy. In both constructing a superior HAR version and picking helpful values of parameters for sharpening powerful test data, the dividing and conquering 1D CNN approach has altered considerably. Their method is straightforward and effective, and it is straightforward to implement once abstract tasks suited for the first level can be identified.

In paper [13], the authors built a complete system for detecting human activity using wearable sensor data. The Adaboost ensemble's classifier is used to interpret human movement. Ensemble classifiers use a weighted blend of a few different types of classifiers to improve performance. The trial results demonstrated the potential of Adaboost ensemble classifiers to provide improved output for digital human movement detection by employing human body sensors. In this framework, the Adaboost ensemble approach is used to obtain high-order precision in order to identify human activity. In this regard, a novel technique has been presented. The proposed model was evaluated for seven distinct physical activities, and general accuracy of 99.98 percent was achieved using Adaboost with random forest. Adaboost-dependent ensemble classifiers significantly enhance the presentation of automated human activity identification, according to a study (HAR).

J. Xie and M.S. Beigi introduced a time-scale invariant feature descriptor for 1D sensor signals in [14]. They created a scale-invariant classifier known as SIC-R. They claimed that their descriptor required fewer training samples since the features are time-scale invariant, hence the speed at which an event happened is irrelevant. They tested it on a variety of sensors, including accelerometers, passive infrared sensors, and seismic 1D sensors, and attained over 90% accuracy.

The authors of [15] developed a Deep Neural Network based on several forms of sensor data to recognize human activities. They created a fusion residual network using computer vision algorithms and image-time series. The rationale for the development of the fusion residual network was the large amount of data acquired from various types of sensors. This design was only tested on two datasets, the HHAR dataset, and the MHEALTH dataset. They employed ResNet to address heterogeneity in the dataset and produced better accuracy results than other strategies.

The authors of [16] looked at the time-window size used in cyclic rehabilitative motion detection to see if it may increase the precision of the Convolutional Neural Network. They employed a three-piece system consisting of a smartwatch, a smartphone, and a server. Subjects were given smartwatches to wear on their wrists for data gathering purposes. The information gathered was then classified. They thought the precision was optimal when the time-window length was equal to or double the sample measuring period. Their work is a terrific initial step, but they only employed one healthy person and just five rehabilitation movements.

The authors of the article [17] discuss the relevance of Human Activity Recognition (HAR) in the treatment of patients and the elderly. They emphasized the advantages of HAR when used in conjunction with machine learning approaches using smart devices and sensors. Smartphones, sensors, and pictures may all be used to collect data.

The study outlines the significance of user behavior and activity recognition (AR) machine learning techniques. Machine learning algorithms and self-evolving feature-based deep learning algorithms can be used to track the techniques on the operation dataset. These techniques help AR in different domains, including personal healthcare, the monitoring of elderly people, functional and behavioral health monitoring using wearables [18].

Using inertial sensors, the researcher [19] provided a thorough description of each phase of the approach typically used to identify human sports with cellphones. The major works from the literature, as well as recommendations for best practices, are presented in the depiction of the steps. Issues relating to the capabilities utilized in categorization models were addressed in particular. In this context, he discusses feature extraction techniques that are entirely dependent on how functions are extracted, i.e., whether they are extracted manually or automatically.

The author of [20] provided a detailed assessment of recent approaches for physiologically stimulated activity. Some new fields were added to the mix, along with a focus on identifying biological activities. Other aspects in the field of machine motion, data processing, psychological and neurological procedures have also been detailed from this perspective in terms of the formulation of guidelines for destiny research.

In [21] the authors provided RF-ARP, which is a 3-stage framework to cope with the problem of HAR in smart homes. Their framework can infer the high-stage activities and also expect the following feasible activity. Compared with the present work, their framework performs better.

The study [22] conducted a study of 32 research papers published between 2011 and 2014. Several HAR sensing technologies were addressed in this research. It demonstrates that RGB cameras have suffered as a result of HAR research when compared to depth sensors and wearable technologies.

In [23] study, researchers recommended the use of wearable instrument recordings called ARN, a specific, asymmetrical residual network for event recognition. In the process, there are two options. One method uses a long frame to capture accurate temporal functions. The alternate method uses a short time interval to capture spatial functions. Extensive research on the two benchmark physical health databases, on the other hand, shows that the ARN outperforms the baselines.

## III. DATASET USED

The dataset [24] we utilized comprises information from inertial sensors on a smartphone called Galaxy SII that was put on the waists of volunteers, including an accelerometer and gyroscopic sensors. The data collecting project enlisted the help of 30 participants, ranging in age from 19 to 48. The details of Classes and their distribution is given in the histogram given (table 1).

TABLE I. SMARTPHONES DATASET ACTIVITIES DETAILS

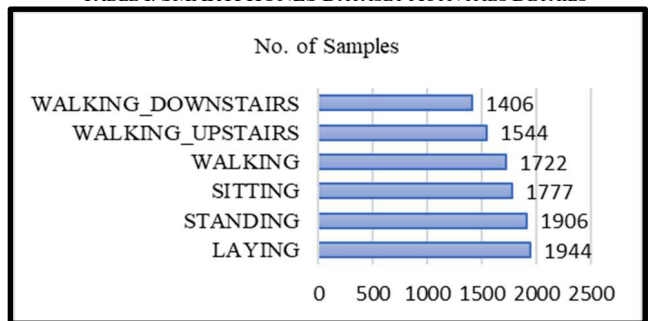

The team has already done some Feature Engineering and have created a feature vector having 561 features from both frequency and time domain. The dataset contains raw sensor data and engineered features data as well.

## IV. MACHINE LEARNING TECHNIQUES USED

We have used all 561 features, which were already engineered. We have used various machine learning classifiers which includes the following
1. K Nearest Neighbors Classifier
2. Support Vector Machine
    a. Linear Kernel
    b. Sigmoid Kernel
    c. Polynomial Kernel
3. Naive Bayesian Classifier
4. Multi-Layer Perceptron Classifier

Sklearn was used for implementing all the machine learning classifiers. The original dataset was divided into 70% Training and 30% Test Data. We further divided 30% of test data into 2 chunks of 15 percent Validation and 15% testing data in order to avoid overfitting. Hyper-parameters were trained on Validation data. Accuracy was considered the measure of performance.

### A. K Nearest Neighbors Classifier

The value of K was increased from 1 to 9. For each value of K accuracy on Validation Data was found. It was observed that accuracy does not increase significantly after K = 5. which was then selected as the final parameter. The accuracy achieved on Test data was 90.43 %. Table 2 below shows the Validation Data Accuracy versus K values

TABLE II. KNN WITH DIFFERENT N VALUES

| K Value | Validation Data Accuracy |
|---|---|
| 1 | 87.78 % |
| 2 | 86.12 % |
| 3 | 89.00 % |
| 4 | 89.13 % |
| 5 | 89.88 % |
| 6 | 90.32 % |
| 7 | 90.76 % |
| 8 | 91.3 % |
| 9 | 91.03 % |

The data in the above table is plotted in a graph shown below (figure 1)

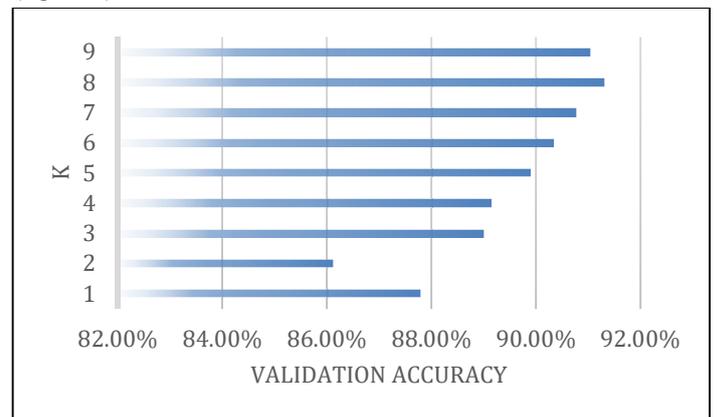

Fig 1: Validation Data Accuracy versus K value

Figure 2 below shows the confusion matrix for KNN Classifier.

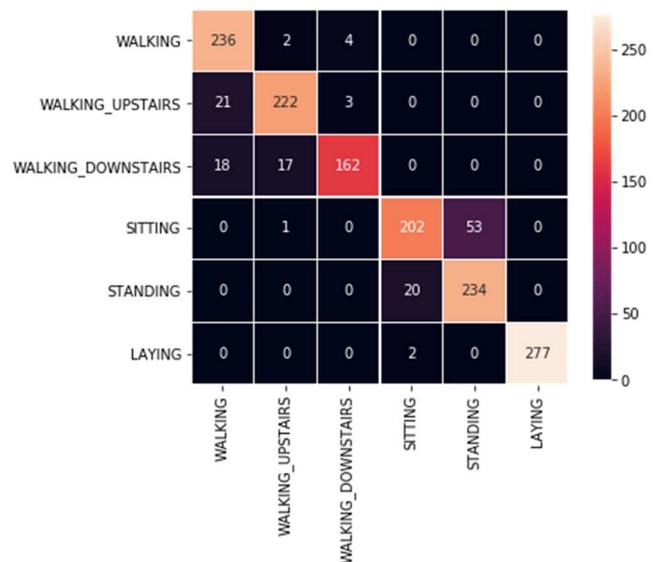

Fig 2: Confusion Matrix of KNN

## B. Support Vector Machine

Three different kernels were used. Namely Linear, Sigmoid, and Polynomial. Regularization factor " C" was set to 0.5. Table 2 below shows the test data accuracy for all SVMs with different kernels.

TABLE II. SVMs WITH DIFFERENT KERNELS

| Kernel | Accuracies |
|---|---|
| Linear | 96.26 % |
| Sigmoid | 91.75 % |
| Polynomial | 90.12 % |

Figures 3, 4 and 5 show the confusion matrices for SVM with Linear, Sigmoid and Polynomial Kernels Respectively.

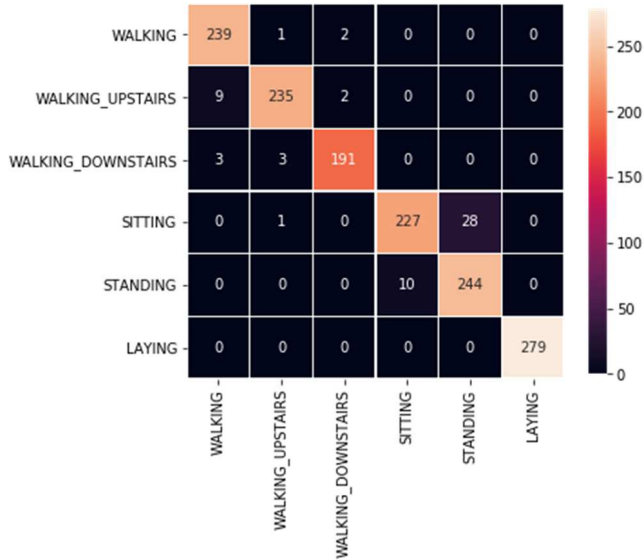

Fig 3: Confusion Matrix SVM Linear Kernel

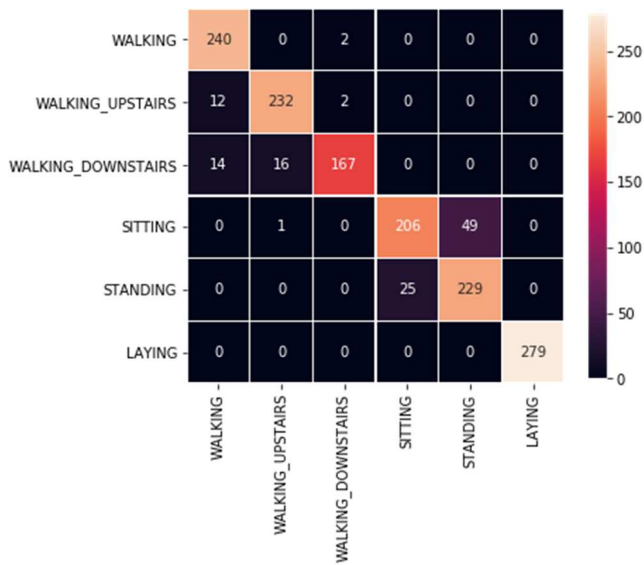

Fig 4: Confusion Matrix SVM Sigmoid Kernel

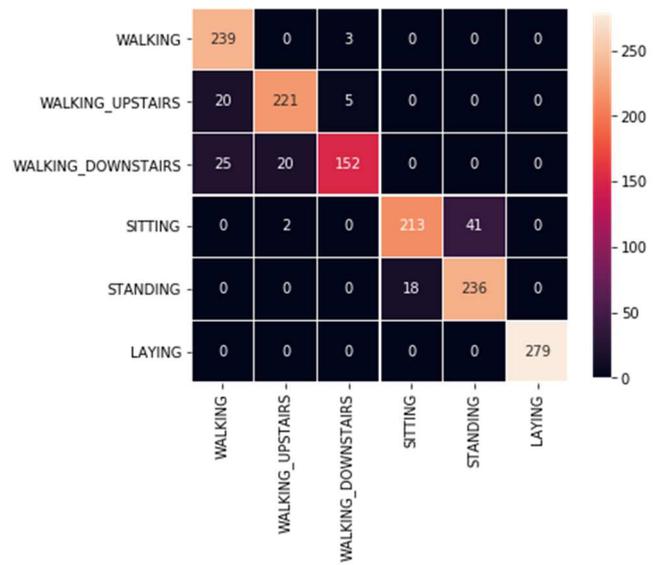

Fig 5: Confusion Matrix SVM Polynomial Kernel

## C. Naive Bayesian Classifier

We have also experimented with a classifier as simple as Naive Bayesian. The accuracy achieved on Validation Data was 77.25 % and accuracy achieved on Test Data 77.02 %. Figure 6 below shows the confusion matrix of the Naive Bayesian Classifier.

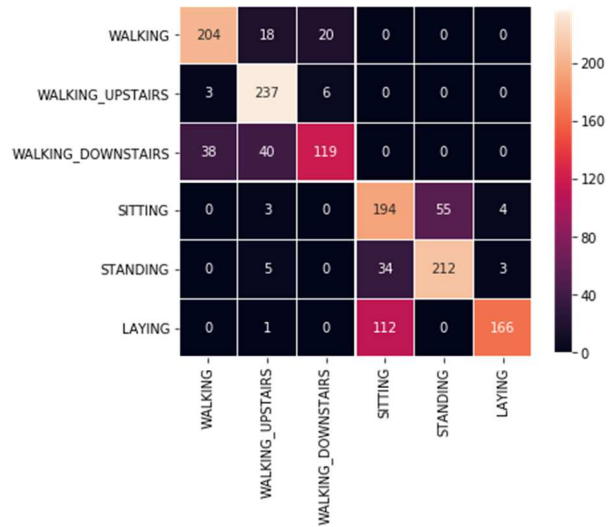

Fig 6: Confusion Matrix Naive Bayesian Classifier

## D. Multilayer Perceptron / Artificial Neural Network

We started with 1 hidden layer and kept on increasing the number of neurons in it. Then increased the number of layers and the same process was repeated. Validation data accuracy was being measured. The best result which we could achieve was on 2 hidden layers with 100 and 65 neurons for the first and second layer respectively. The learning rate was set to 0.001. Total epochs were 1000 and the Relu activation function was being used. Test data accuracy achieved was 94.23 %. Figure 7 below shows the confusion matrix for ANN training.

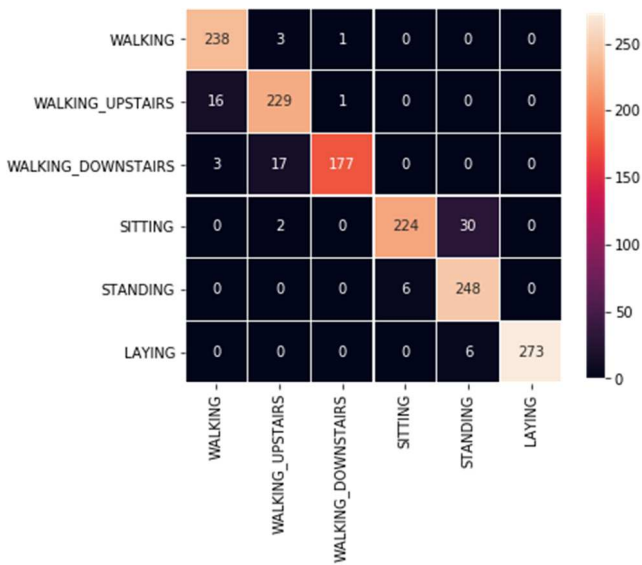

Fig 7: Confusion Matrix MLP Classifier

## V. COMPARISON OF RESULTS

We have implemented different Machine Learning classifiers and trained them on the feature engineered data table 3 summarizes the test data accuracy for all the classifiers.

TABLE III. COMPARISON OF RESULTS

| Classifiers | Test Data Accuracy % |
| --- | --- |
| **K Nearest Neighbors Classifiers** | 90.43 |
| **Multi-layer Perceptron** | 94.23 |
| **Naïve Bayesian Classifier** | 77.02 |
| **SVM with Linear Kernel** | 96.26 |
| **SVM with Polynomial Kernel** | 90.12 |
| **SVM with Sigmoid Kernel** | 91.75 |

This conclusion is SVM with linear kernel performs the best with the dataset. Achieving accuracy upto 96.26 percent.

## VI. FUTURE DIRECTIONS

This research is part of our initial work on HAR. We are currently working on building deep neural networks that employ raw sensor data from inertial sensors in wearable devices to classify human activities. Another dimension in which our team is working is converting signals into images using various transformation techniques in order to transfer the domain to image classification, which is one of the most active areas in the domain of deep and machine learning.

## VII. CONCLUSIONS

Human Activity Recognition is a topic of current research. The use of sensor data for HAR is becoming famous and a lot of effort has been made in using different deep learning and machine learning techniques for HAR. We have implemented different ML and DL techniques and have done their comparison. Results show that Support Vector Machine with linear Kernel performs the best for such applications. Hence instead of using complex deep learning architectures one should spend time in feature engineering and should use simple machine learning classifiers.